\documentstyle[prl,aps,twocolumn,psfig]{revtex}

\begin{document}
\twocolumn[\hsize\textwidth\columnwidth\hsize\csname
@twocolumnfalse\endcsname

\draft

\title{Narrow Spectral Response of a Brillouin Amplifier}
\author{Shmuel Sternklar and Er'el Granot}

\address{Department of Engineering, Academic College of Judea and Samaria, Ariel 3,
Israel}
\date{\today}
\maketitle

\begin{abstract}
\begin{quote}
\parbox{16 cm}{\small
We investigate the spectral response of a Brillouin amplifier in
the frequency regime within the SBS bandwidth. This is done by
\textit{amplitude }modulating the pump with a low frequency, and
therefore, unlike previous studies, the spectrum of the modulated
pump is, in all cases, smaller than the SBS bandwidth. We show
both theoretically and experimentally that unlike phase
modulation, which was reported in the literature, the amplitude
modulation \textit{increases} the Brillouin amplifier gain, and
that this effect has a very narrow bandwidth. Only modulation
frequencies that are lower than a certain cut-off frequency
increase the gain. This cut-off frequency is inversely
proportional to the fiber's length, and can therefore be
arbitrarily small. }
\end{quote}
\end{abstract}

 ]

\narrowtext \noindent

In long-distance optical communications in fibers the first
nonlinear effect to appear is stimulated Brillouin scattering
(SBS). Many investigations over the past decade have studied the
influence of SBS on data transmission, either to evaluate the
induced degradation of the optical signal or to utilize it for
optical processing
\cite{1_Chen_Bao,2_Zho_Shalaby_Chao_Cheng_Ye,3_Esman_Williams,4_Loayssa_Benito_Garde}.
Clearly, the optical signal quality may be severely deteriorated
due to SBS, since the signal experiences significant depletion
over long distances. However, only a portion of the signal's
spectrum is depleted. It is well known, that the Brillouin process
is characterized by a finite spectral width, $\Gamma _B $, which
is related to the phonon lifetime in the material (in glass
fibers, for example, $\Gamma _B / 2\pi \cong 20GHz)$. Hence, every
spectral component beyond the Brillouin spectral width will not
take part in the Brillouin process, and will not be depleted.

Due to the practical importance of Brillouin scattering in long
distance fiber communications, and its negative effect on the
transmitted optical signal, most of the studies investigated the
possibility of suppressing SBS. It was determined that SBS could
be almost entirely suppressed by modulating the incident optical
signal with a frequency that exceeds tens of MHz
\cite{5_Bolle_Grosso_Daino,6_Tsubokawa_Seikai_Nakashima_Shibata}.
Due to the spectral width of the SBS, this is not a surprising
finding.

Other works \cite{7_Hook_Bolle_Grosso_Martinelli,8_Hook_Bolle},
which investigated the temporal response to lower frequencies,
usually employed square-wave modulation (having a fast rise time),
so that the lower portion of the spectrum did not received enough
attention. There were however, some indications
\cite{9_Eskildsen_et_al} that the Brillouin amplifier can react in
a non-trivial manner to modulation frequencies, which are
considerably lower than the SBS's spectral width. Eskildsen and co
workers \cite{9_Eskildsen_et_al} showed that SBS can be
substantially suppressed by directly modulating a DFB laser with a
frequency as low as 5kHz. However, in their experiment, the
external low frequency modulation caused an extremely large (much
larger than the SBS spectral width) wavelength dithering, i.e., a
much larger broadening.

In this paper we investigate the spectral response of the
Brillouin amplifier to modulation frequencies lower than the SBS
bandwidth. This is done by \textit{amplitude }modulating the pump
with a low frequency, and therefore, unlike previous work
\cite{5_Bolle_Grosso_Daino,6_Tsubokawa_Seikai_Nakashima_Shibata,9_Eskildsen_et_al},
the spectrum of the modulated pump is, in all cases, smaller than
the SBS bandwidth. We show both theoretically and experimentally
that unlike phase modulation, which was reported in the
literature, the amplitude modulation \textit{increases} the
Brillouin amplifier gain, and that this effect has a very narrow
bandwidth. Only modulation frequencies, which are lower than a
certain cut-off frequency increases the gain. This cut-off
frequency value $f_c = c / 2nL$ , which depends only on the
fiber's length ($L)$ and can be much smaller than the Brillouin
spectral width.

In the slowly varying amplitude approximation \cite{10_Boyd}

\begin{equation}
\label{eq1}
\frac{\partial A_2 }{\partial z} + \frac{1}{c / n}\frac{\partial A_2
}{\partial t} = \frac{i\omega \gamma _e }{2nc\rho _0 }\rho ^\ast A_1
\end{equation}

\noindent
where

$A_1 $and $A_2 $ are the Stokes and pump amplitudes respectively, $n$ is the
refractive index, and

\begin{equation}
\label{eq2}
\rho \left( {z,t} \right) = i\frac{\gamma _e q^2}{4\pi }\frac{A_1 A_2^\ast
}{\Omega _B \Gamma _B }
\end{equation}

\noindent
is the material density distribution,

\noindent where $\gamma _e $ is the electrostrictive constant,
$\Gamma _B $ and $\Omega _B $ are the Brillouin linewidth and
frequency respectively, $q$ is the acoustic wave number, and
$\omega$ is the optical angular frequency .

Eq.(\ref{eq1}) can be rewritten in terms of the intensities
\cite{10_Boyd}

\begin{equation}
\label{eq3}
\frac{\partial I_2 }{\partial z} + \frac{1}{c / n}\frac{\partial I_2
}{\partial t} = g_0 I_1 \left( {z,t} \right)I_2
\end{equation}

\noindent
where the line center gain factor is

\begin{equation}
\label{eq4}
g_0 \equiv \frac{\omega ^2\gamma _e^2 }{2nc^3v\rho _0 \Gamma _B },
\end{equation}

$v$ is the acoustic velocity and $\rho _0 $ is the mean density of
the medium.

In the non-depleted pump approximation we can describe the modulated pump as

\begin{equation}
\label{eq5} I_1 \left( {z,t} \right) = I_1(L)\left\{ {1 + \cos
\left[ {2\pi f\left( {zn / c + t} \right)} \right]} \right\}
\end{equation}

\noindent and then the solution to eq.\ref{eq3} for the amplified
Stokes will satisfy

\begin{equation}
\label{eq6}
\begin{array}{l}
 \ln \left[ {I_2 \left( {z,t} \right)} \right] \propto \\
 C_1 \sin \left[ {2\pi f\left( {zn / c + t} \right)} \right] + C_2 \sin
\left[ {2\pi f\left( {zn / c - t} \right)} \right] + \\
C_3 \cos\left[ {2\pi f\left( {zn / c - t} \right)} \right] + \\
 C_4 \left( {zn / c + t} \right) + C_5 \left( {zn / c - t} \right) \\
 \end{array}
\end{equation}

(the first and the fourth terms correspond to the specific
solution while the other terms correspond to the homogenous ones),
where due to the boundary condition [$I_2 \left( {z = 0} \right) =
I_2 \left( 0 \right)$], $C_2 = C_1 $ ,$C_3 = 0$ and $C_4 = C_5 $,
and the final solution at $z = L$ is

\begin{equation}
\label{eq7} I_2 (L) = I_2(0)\exp \left\{ {G\left[ {1 + \cos \left(
{2\pi ft} \right)\frac{\sin \left( {2\pi fnL / c} \right)}{2\pi
fnL / c}} \right]} \right\}
\end{equation}

\noindent where $G \equiv g_0 I_1(0)L$ (in our experiment $G \cong
1)$.

This solution oscillates in time. The difference between the maximum and
minimum intensities corresponding to the oscillatory portion of the signal
is

\begin{equation}
\label{eq8} \Delta I_2 = 2I_2(0)\exp \left( G \right)\sinh \left[
{G\left| {\frac{\sin \left( {\pi f / f_c } \right)}{\pi f / f_c }}
\right|} \right]
\end{equation}

\noindent
where

\begin{equation}
\label{eq9}
f_c \equiv c / 2nL
\end{equation}

\noindent
is the cut-off frequency (the first frequency where $\Delta I_2 = 0)$

This solution suggests that the Brillouin effect in the fiber reacts to
frequencies, which depend solely on the fiber's length. For long fibers,
these frequencies can be considerably smaller than the Brillouin spectral
width.

It should be noted, that this effect is not necessarily specific to the
Brillouin process. Other nonlinear, distributed optical amplifiers, when
pumped by an amplitude-modulated beam, will present a similar spectral
response. The reasoning behind this is as follows. Since the amplification
is a nonlinear function of the pump intensity, the impact of the temporal
maxima and minima of the modulated pump do not cancel each other; the maxima
have an excess gain (their gain increase is larger than the gain decrease at
the minima), and therefore the Stokes beam experiences a net increase in
gain. However, when the fiber length is larger than the modulation
wavelength the effective gain averaged over the spatial maxima and minima
converges to the unmodulated gain value.

\begin{figure}
\psfig{figure=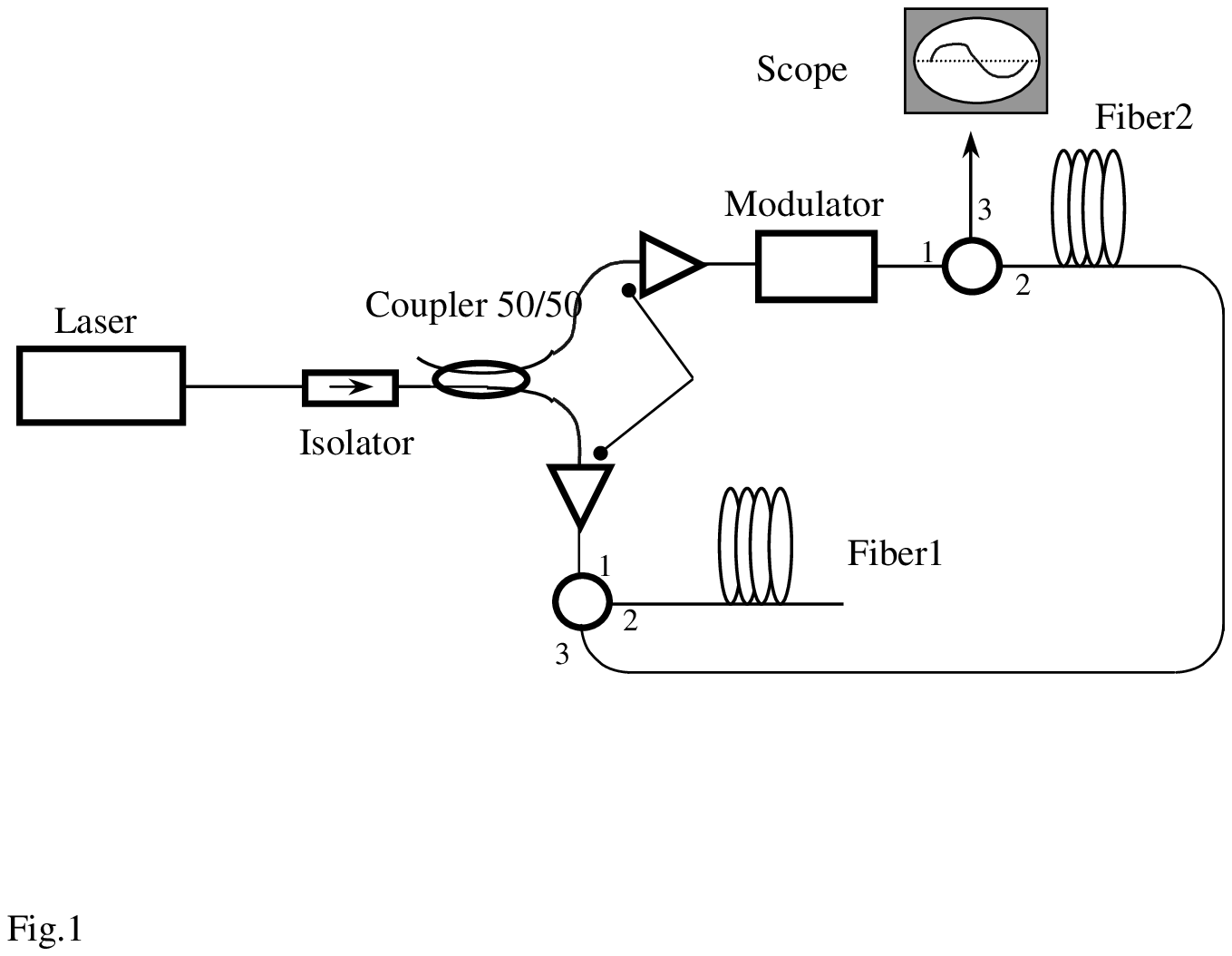,width=10cm,bbllx=90bp,bblly=500bp,bburx=550bp,bbury=750bp,clip=}
\caption{Illustration of the system.}\label{fig1}
\end{figure}

The experimental set-up is shown in Fig. 1. Light from a narrow-band ($\sim
$200kHz) laser (NetTest Tunics Plus) at 1550nm is split by a coupler. A
portion is amplified by an EDFA and sent into Fiber 1, to generate a
Brillouin Stokes beam with a bandwidth of about 30Mhz, which is then guided
into Fiber 2 as the seed for the Brillouin amplifier. The other portion is
also amplified and then modulated by a LiNbO3 electro-optic modulator,
before being sent into the other side of Fiber 2, to act as the pump for the
Brillouin amplifier. The amplified Stokes exits through the circulator.

\begin{figure}
\psfig{figure=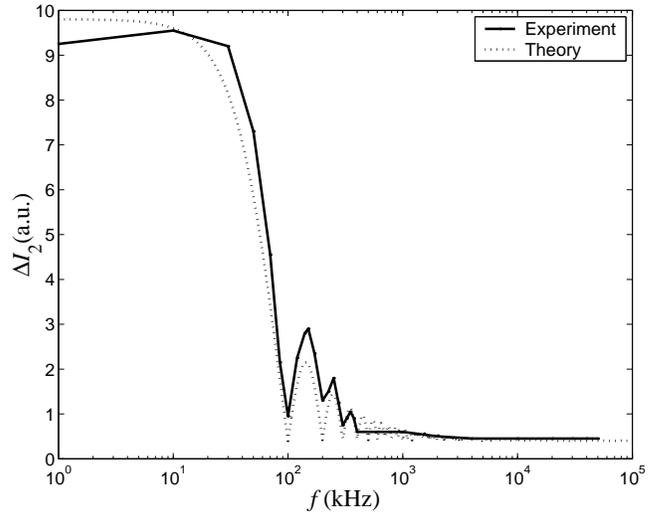,width=10cm,bbllx=35bp,bblly=1bp,bburx=620bp,bbury=420bp,clip=}
\caption{The frequency response of the AC part of the amplified
Stokes signal, where Fiber2 was 1km long.}\label{fig2}
\end{figure}

We investigated two different lengths of single-mode fiber for the
Brillouin amplifier: $ \sim $1km and $ \sim $2km. In both cases,
the pump power entering Fiber 2 was approximately 10mW, and the
Stokes power was about three orders of magnitude lower. The
amplified Stokes consisted of an amplified dc component as well as
an ac component due to the presence of the modulated pump. We
measured the peak-to-peak amplitude of the amplified Stokes as a
function of the modulation frequency. The experimental results and
theoretical prediction are shown in Figs. 2 and 3.

\begin{figure}
\psfig{figure=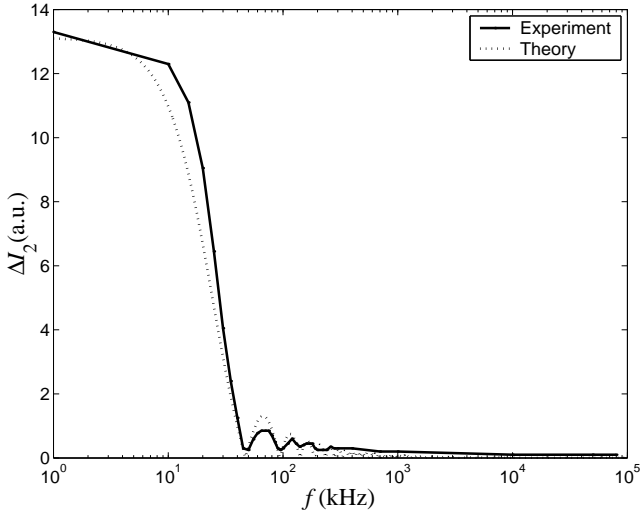,width=10cm,bbllx=35bp,bblly=1bp,bburx=620bp,bbury=420bp,clip=}
\caption{The frequency response of the AC part of the amplified
Stokes signal, where Fiber2 was 2km long.}\label{fig3}
\end{figure}

For the 1km amplifier, the cut-off frequency is 100 kHz, and for
the 2km amplifier it is 50kHz, in perfect agreement with theory.

In general, where the pump in eq.\ref{eq5} has a general form $I_1
\left( {z,t} \right)$ it can always be separated into its Fourier
components $I_1 \left( {z,t} \right) = I_1\left( L \right)\int
{dfa_f \cos \left[ {2\pi f\left( {zn / c + t} \right)} \right]}$.
Clearly, a generalization of \ref{eq6} will still be valid, and,
as a consequence, only the low frequencies components will
contribute to the amplification. We therefore conclude that the
Brillouin amplifier can behave like a narrow band amplifier, whose
cut-off frequency is inversely proportional to its length, and can
be considerably narrower than the Brillouin spectral width.

\end{document}